\documentclass[twocolumn,prc,floatfix,showpacs,preprintnumbers,nofootinbib,%
superscriptaddress]{revtex4}
\usepackage{mathrsfs}
\usepackage{amssymb}
\usepackage{amsmath}
\usepackage{graphicx}
\usepackage{hyperref}
\usepackage{mciteplus}

\newcommand{\rt}{{\mathbf{r}_T}}

\newcommand{\bt}{{\mathbf{b}_T}}

\newcommand{\Deltat}{{\boldsymbol{\Delta}_T}}

\newcommand{\ud}{\, \mathrm{d}}

\newcommand{\nc}{{N_\mathrm{c}}}

\newcommand{\nr}[1]{(\ref{#1})}

\newcommand{\ra}{R_A}
\newcommand{\rp}{R_p}

\newcommand{\gev}{\ \textrm{GeV}}
\newcommand{\fm}{\ \textrm{fm}}

\newcommand{\qs}{Q_\mathrm{s}}

\newcommand{\as}{\alpha_{\mathrm{s}}}

\newcommand{\fig}{Fig.~}
\newcommand{\figs}{Figs.~}
\newcommand{\eq}{Eq.~}

\newcommand{\eqs}{Eqs.~}

\newcommand{\sigmap}{{ \sigma^\textrm{p}_\textrm{dip} }}

\newcommand{\dsigmap}{{\frac{\ud \sigma^\textrm{p}_\textrm{dip}}{\ud^2 \bt}}}

\newcommand{\dsigma}{{\frac{\ud \sigma_\textrm{dip}}{\ud^2 \bt}}}

\newcommand{\xpom}{{x_\mathbb{P}}}

\newcommand{\Aavg}[1]{\left\langle #1 \right\rangle_\textrm{N}}

\newcommand{\ampli}{{\mathcal{N}}}
\newcommand{\A}{{\mathcal{A}}}

\begin{document}

\author{T. Lappi}
\affiliation{
Department of Physics, %
 P.O. Box 35, 40014 University of Jyv\"askyl\"a, Finland
}

\affiliation{
Helsinki Institute of Physics, P.O. Box 64, 00014 University of Helsinki,
Finland
}
\author{H. M\"antysaari}
\affiliation{
Department of Physics, %
 P.O. Box 35, 40014 University of Jyv\"askyl\"a, Finland
}

\title{
Incoherent diffractive $J/\Psi$-production in high energy nuclear DIS
}

\pacs{13.60.Hb,24.85.+p}

\preprint{INT-PUB-10-061}

\begin{abstract}
We compute cross sections for incoherent diffractive $J/\Psi$ production
in lepton-nucleus deep inelastic scattering. The cross section
is proportional to $A$ in the dilute limit and to $A^{1/3}$ in the
black disc limit, with a large nuclear suppression due to saturation effects.
The $t$-dependence of the cross section, if it can be measured accurately enough,
is sensitive to the impact parameter
profile of the gluons in the nucleus and their fluctuations, a quantity
that determines the initial conditions of a relativistic heavy ion collision.
The nuclear suppression in incoherent diffraction shows  how
the transverse spatial distribution of the gluons in the nucleus
gradually becomes smoother at high energy.
Since the values of the momentum transfer $|t|$ involved are 
relatively large, this process should be easier to measure 
in future nuclear DIS  experiments than coherent diffraction. 
\end{abstract}

\maketitle

\section{Introduction}

Strongly interacting systems in the high energy (or small $x$) limit
are very nonlinear systems in spite of the smallness of
the coupling constant $\as$. This is due to the large phase
space available for semihard gluon radiation that increases 
the occupation numbers of gluonic modes in the hadron or nucleus 
wavefunction. Thus high energy scattering has to be understood in
terms of gluon recombination and saturation that enforce the unitarity
requirements of the $S$-matrix. This happens naturally in
the Color Glass Condensate (CGC) effective theory of the
high energy wavefunction. In the context of 
deep inelastic scattering (DIS) the CGC 
leads to the dipole picture that naturally
gives a consistent description of both inclusive
and diffractive scattering.
The nonlinearities in high energy scattering are enhanced when the target
is changed from a proton to a heavy nucleus. Thus there is a great
opportunity to understand them by studying nuclear DIS in 
new collider experiments, such as the EIC~\cite{Deshpande:2005wd} 
or the LHeC~\cite{Dainton:2006wd}. The particular process we discuss in this
paper is diffractive DIS on nuclei.

In the Good-Walker~\cite{Good:1960ba} picture of diffraction one needs to identify 
the states
that diagonalize the imaginary part of the $T$-matrix. In the case of 
nuclear DIS at high energy these states are the ones with the virtual photon 
fluctuating into a dipole of a fixed size $r$ and with the nucleons in
the nucleus at fixed transverse positions $b_i$. In coherent diffraction
the nucleus is required to stay intact, which corresponds to performing
 the average over the nuclear wavefunction at the level of the
scattering amplitude. 
Averaging the cross section, instead of the amplitude,  
over the nucleon positions allows for the nucleus
to break up,  giving the sum of incoherent and coherent cross sections,
i.e. the quasielastic cross section. 
For a  more formal discussion of this we point the reader e.g. to 
Ref.~\cite{Caldwell:2009ke}.
The $t$-dependence of the incoherent cross section therefore directly 
probes the fluctuations and correlations in the nuclear wavefunction, which
have turned out to be a crucial ingredient in understanding the initial 
conditions in heavy ion collisions \cite{Miller:2007ri,*Alver:2010gr}.

The average gluon density probed in the 
coherent process is very smooth, meaning that the cross section is dominated
by small values of momentum transfer to the nucleus, $t \sim - 1/\ra^2$.
Measuring such a small momentum transfer accurately is 
very challenging.
 At momentum scales corresponding to the nucleon 
size $t \sim - 1/\rp^2$ the diffractive cross section is almost purely incoherent.
The larger momentum transfer should
also be easier to reconstruct experimentally even without measuring
the transverse momentum of the 
nuclear remnants, by accurately reconstructing the outgoing electron
and $J/\Psi$ momenta and using momentum conservation.
By taking these processes into account in the detector design
one should be capable of measuring diffractive events at a higher accuracy
than was done at HERA.
In the dilute limit (for small dipoles) there is no multiple scattering, and the
incoherent cross section is given by $A$ times the corresponding one for
protons.  The
deviation of the $t$-slope from the proton measures the transverse size of the 
fluctuating areas in the nucleus.

In the black disc limit 
the nucleus is smooth not only on average, but event-by-event, 
leading to a strong suppression of the incoherent cross section. Incoherent
diffraction gets contributions from the edge of the nucleus, making the
cross section asymptotically behave as $\sim A^{1/3}$ in contrast to
$\sim A$ in the dilute limit.
The suppression in the
normalization relative to the proton is a measure of the approach to the unitarity
limit in the dipole cross section. It is a clear signal of how individual nucleons
have lost their identity in the sense that they cannot be resolved by the virtual photon.
It is precisely this suppression that we are proposing to 
use to quantitatively access saturation effects in the nuclear  wavefunction.
The purpose of this paper  is to provide a realistic estimate of 
the nuclear suppression in diffractive
cross sections in a  regime that could be measured 
in future nuclear DIS experiments.

Nuclear DIS data from fixed target experiments, in particular 
E665~\cite{Adams:1994bw} and NMC~\cite{Arneodo:1994qb,*Arneodo:1994id} 
have already been much discussed in the literature as
demonstrations of \emph{color transparency} (see e.g. 
Refs.~\cite{Frankfurt:1991nx,Frankfurt:1993it,Brodsky:1994kf,Kopeliovich:2001xj,Frankfurt:2005mc}). The form of nuclear modification to the incoherent
diffraction in terms of the dipole cross section 
that we have rederived is not new (see 
e.g.~\cite{Kopeliovich:1991pu,Kopeliovich:2001xj}).
So far, however, less  attention
has been paid to inelastic diffraction in future DIS experiments.
The production 
cross sections have not been calculated using the same
CGC inspired cross sections that have been used
successfully to confront HERA data, as we intend to do here.
In this work we concentrate on the $J/\Psi$ 
because its small size means that the interaction
of the dipole with the target is calculable in weak coupling even at
small $Q^2$.

The importance of diffraction in understanding gluon saturation has been discussed
and our basic setup motivated in Ref.~\cite{Kowalski:2007rw}.
Nuclear modifications 
to the diffractive structure functions, integrated over the momentum transfer
$t$, were computed in Ref.~\cite{Kowalski:2008sa}. Vector meson production 
at future DIS experiments was recently 
discussed from a more experimental point of view in
Ref.~\cite{Caldwell:2009ke}, and coherent production cross sections (integrated
over $t$) calculated in Ref.~\cite{Goncalves:2009za}.
An interesting discussion on coherent and incoherent diffraction
and gluon saturation in the nucleus can be found in Ref.~\cite{Tuchin:2008np}.
In this study we want to take a step beyond the discussion 
of inclusive diffraction in Refs.~\cite{Kowalski:2007rw,Kowalski:2008sa}
 to understand the $t$ dependence in more detail.

\section{Dipole cross sections}
\label{sec:dipxs}

There are many dipole cross section parametrizations available 
in the literature, and
we have taken for this study two representative samples. One is the 
IIM~\cite{Iancu:2003ge} dipole cross section, which is a 
parametrization including the most important features 
of BK~\cite{Balitsky:1995ub,*Kovchegov:1999yj,*Kovchegov:1999ua}
evolution. The detailed expression for the
dipole cross section can be found in Ref.~\cite{Iancu:2003ge};
we use here the values of the parameters from the newer
fit to HERA data including charm~\cite{Soyez:2007kg}
that was also used to compute diffractive structure functions in
Ref.~\cite{Marquet:2007nf}. We also want to compare
our results to a parametrization with an eikonalized DGLAP-evolved gluon distribution.
For this purpose we will use an approximation of the IPsat dipole 
cross section~\cite{Kowalski:2003hm,Kowalski:2006hc}.

To extend the dipole cross section from protons to nuclei
 we will take the independent
scattering approximation that is usually used in Glauber theory 
and write the $S$-matrix as
\begin{equation}\label{eq:sfact}
S_A(\rt,\bt,x) = \prod_{i=1}^A S_p(\rt,\bt-\bt_i,x).
\end{equation}
Here we conventionally parametrize the energy dependence of the scattering
amplitude with $x$, the Bjorken variable of the DIS event\footnote{
Note that strictly speaking the relation between $x$ and the 
energy of the dipole-target scattering depends 
on $Q^2$, not only $r$. Using $x$ here is justified
in a high energy approximation
where the energy of the dipole in the target rest frame
is approximately the same as that of the virtual photon.}.
The variables $\bt_i$  in \eq\nr{eq:sfact} 
are the nucleon coordinates that we will discuss in 
Sec.~\ref{sec:comp}.
This independent scattering assumption is
natural in IPsat-like parametrizations or the MV~\cite{McLerran:1994ni} model, 
where, denoting $r = |\rt|,$ $S(\rt) \sim e^{-r^2 \qs^2/4}$ with a  saturation scale 
$\qs^2$ proportional to the nuclear thickness $T_A(b)$.
High energy evolution, however, introduces an anomalous dimension that leads,
in the nuclear case, to what could be called leading twist shadowing.
With an anomalous dimension
$S\sim e^{-(\qs r)^{2\gamma}}$ with $\gamma \neq 1$, a proportionality
$\qs^2 \sim T_A(b)$ is not equivalent to \eq\nr{eq:sfact}. A solution to this 
problem (see also the more detailed discussion in~\cite{Kowalski:2008sa}) 
would require a realistic impact parameter dependent solution to the 
BK equation which, we feel fair to say, is not yet available.
We point the reader e.g. to Ref.~\cite{GolecBiernat:2003ym} for a 
discussion of the  difficulties. These are related to the long distance 
Coulomb tails that, physically, are regulated at the confinement length
scale that is not enforced in a first principles weak coupling calculation.
The effect of BK evolution is important for the CGC description
of the forward suppression of particle production
in dAu-collisions at RHIC (for a review see~\cite{Jalilian-Marian:2005jf}).
In our case the difficulty is greater since we are interested
not only in the relatively smooth average gluon density, but
its variations at smaller length scales of the order of the
proton radius.
We thus leave the modifications of \eq\nr{eq:sfact} due to
the effects of evolution to a future study.

\begin{figure}
\includegraphics[width=0.5\textwidth]{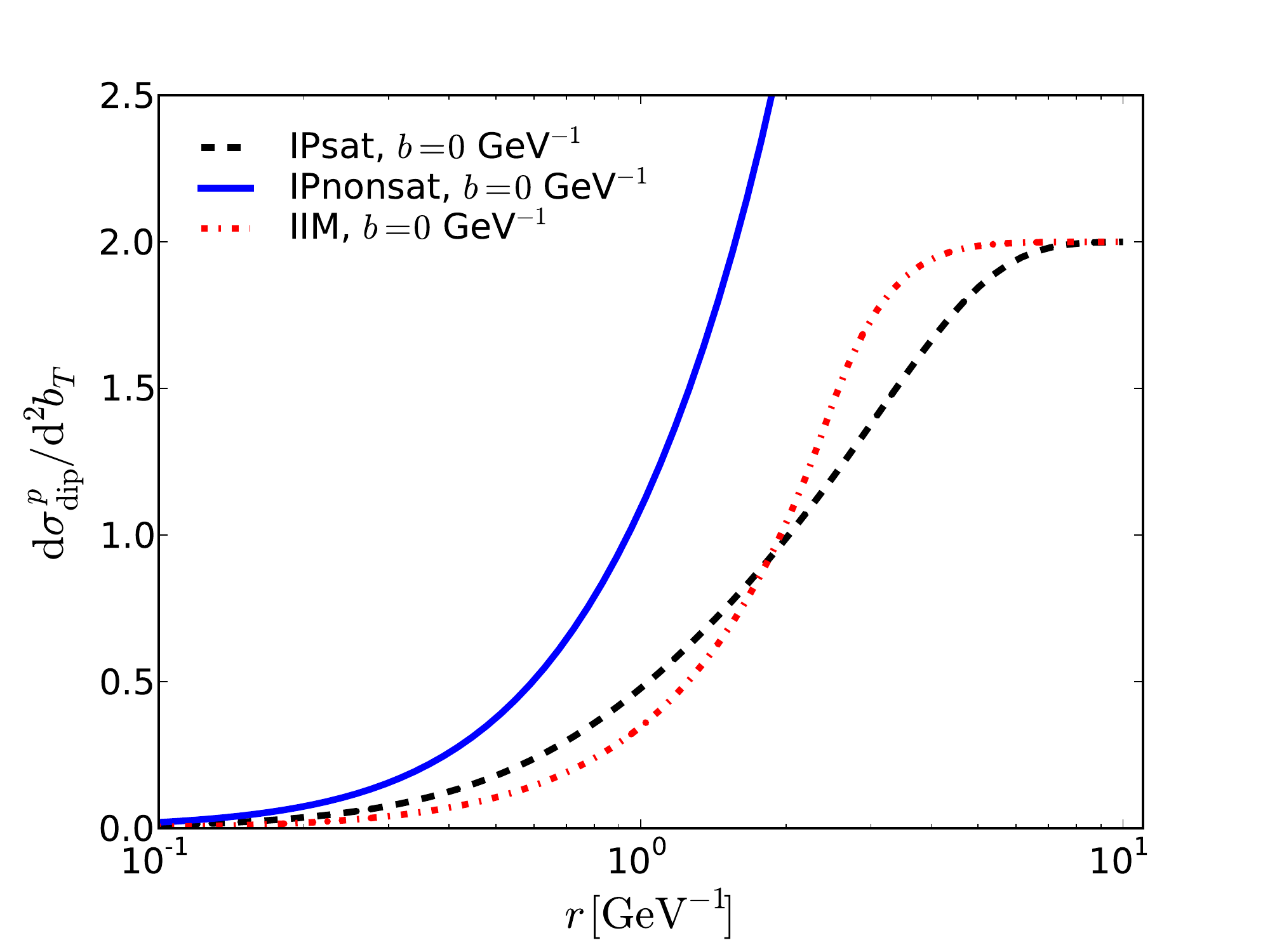}
\caption{
The $r$-dependence of the different proton dipole cross  sections used,
at  $x=0.0001$ and $b=0$.
As discussed in Sec.~\ref{sec:res}, the ``IPnonsat''-curve is
\eq\nr{eq:BEKWfact} linearized in $r^2F(x,r)$.
} \label{fig:sigmap}
\end{figure}

The IIM parametrization assumes, either explicitly or implicitly,
a factorizable $\bt$ dependence 
\begin{eqnarray}\label{eq:factbt}
\dsigmap(\bt,\rt,x) &=&  2 \left( 1 - S_p(\rt,\bt,x)\right)
\\ \nonumber
&=& 2 \,T_p(\bt) \ampli(r,x),
\end{eqnarray}
We take, following Ref.~\cite{Marquet:2007nf}, a Gaussian profile
$T_p(\bt) = \exp\left(-b^2/2 B_p\right)$ with 
$B_p=5.59\gev^{-2}$ (see Sec.~\ref{sec:res} for a discussion of 
this largish numerical value). 

In the IPsat model the impact parameter dependence is
included in the saturation scale as
\begin{equation}\label{eq:unfactbt}
\dsigmap(\bt,\rt,x)
 = 2\,\left[ 1 - \exp\left(- r^2  F(x,r) T_p(\bt)\right) 
\right].
\end{equation}
Here $T_p(\bt) = \exp\left(-b^2/2 B_p\right)$ 
is the impact parameter profile function in the proton 
with $B_p=4.0\gev^2$ and $F$ is proportional to the 
DGLAP evolved gluon distribution~\cite{Bartels:2002cj}
\begin{equation}
F(x,r^2) = 
\frac{1}{2 \pi B_p}
\frac{ \pi^2 }{2 \nc} \as \left(\mu_0^2 + \frac{C}{r^2} \right) 
x g\left(x,\mu_0^2 + \frac{C}{r^2} \right),  
\label{eq:BEKW_F}
\end{equation}
with $C$ chosen as 4 and $\mu_0^2=1.17\gev^2$ resulting from the 
fit~\cite{Kowalski:2006hc}. The proton dipole cross sections used are
plotted in \fig\ref{fig:sigmap}  for $x=0.0001$.

We would generally prefer the unfactorized $b$-dependence
of \eq\nr{eq:unfactbt} to the factorized one in \eq\nr{eq:factbt}
because it allows for the correct unitarity
limit of the scattering amplitude at all impact parameters 
(see the discussion in Ref.~\cite{Kowalski:2008sa}). 
However, there seems to be no clear difference between the two in
terms of the quality of the description of HERA data, and for the sake 
of computational simplicity we will in this work limit ourselves to
the factorized dependence and approximate the IPsat dipole cross section
by
\begin{equation}\label{eq:BEKWfact}
\dsigmap(\bt,\rt,x)
 \approx  2 T_p(\bt) \,\left[ 1 - \exp\left(- r^2  F(x,r)\right)
\right]
\end{equation}
using the same $F(x,r)$ defined in \eq\nr{eq:BEKW_F}. This approximation
brings the IPsat parametrization to the form \eq\nr{eq:factbt}
with  $\ampli(r,x)=\left[ 1 - \exp\left(- r^2  F(x,r)\right)\right]$;
in fact this is the form used already in Ref.~\cite{Bartels:2002cj}; we however
use the gluon distribution from the IPsat fit~\cite{Kowalski:2006hc} 
for convenience. Improving this
description goes hand in hand with giving up the approximation
of independent scatterings off the nucleons, \eq\nr{eq:sfact},
and is left for future work. As we shall see in the following, these 
approximations enable us to write the cross section for incoherent
diffraction in a form which is much simpler to evaluate numerically 
than one with a general $b$-dependence.

\begin{figure}
\includegraphics[width=0.5\textwidth]{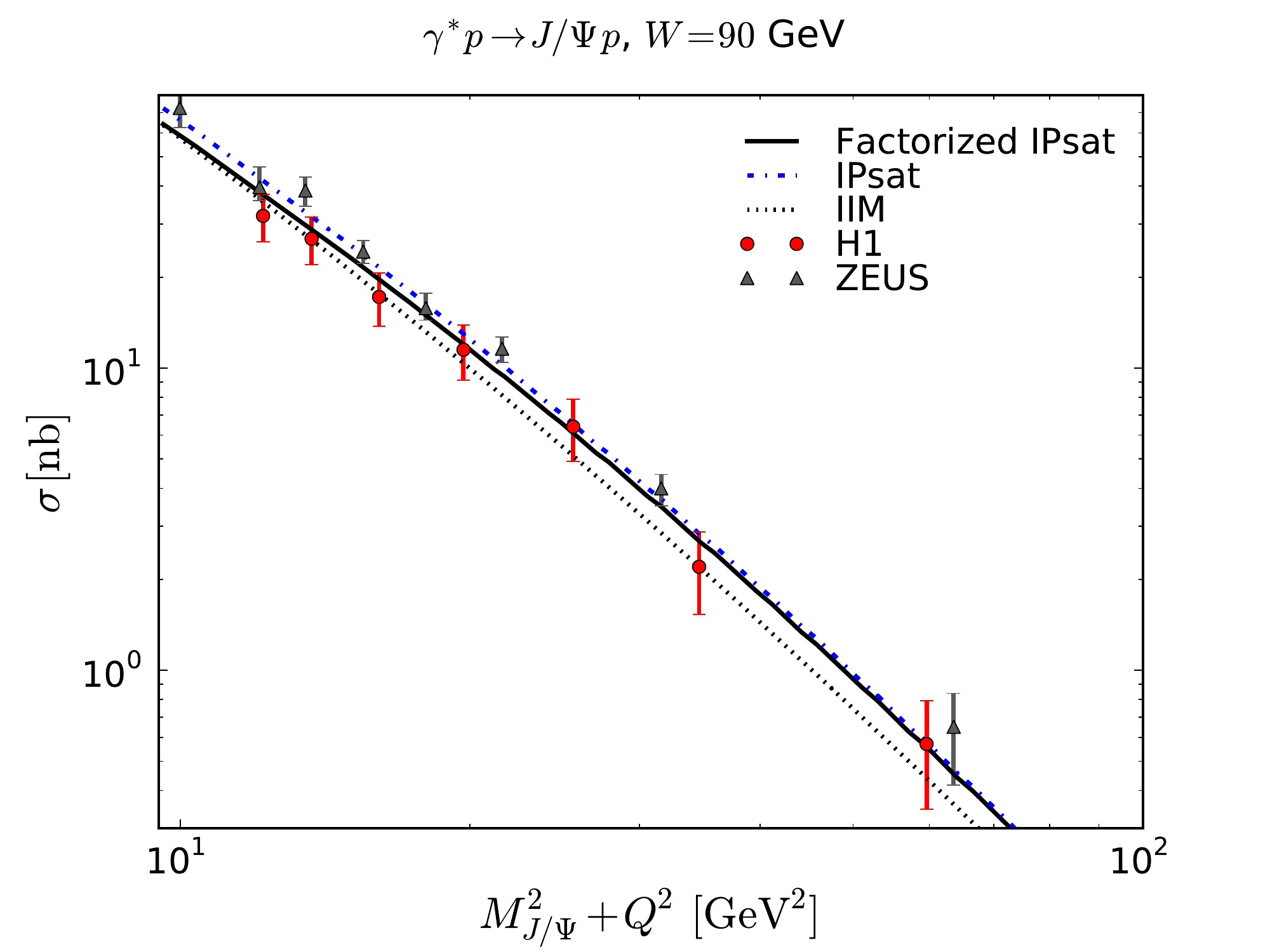}
\caption{Comparison of the used dipole cross sections to HERA 
data~\cite{Chekanov:2004mw,*Aktas:2005xu} on diffractive vector meson production.
} \label{fig:hera}
\end{figure}

\section{Computing diffractive cross sections}
\label{sec:comp}

The cross section for quasielastic vector meson 
production in nuclear DIS is
\begin{equation} \label{eq:xsec}
\frac{\ud \sigma^{\gamma^* A \to V A }}{\ud t} 
= \frac{R_g^2(1+\beta^2)}{16\pi} \Aavg{|\A(\xpom,Q^2,\Deltat)|^2}.
\end{equation}
with $t=-\Deltat^2$. 
The dipole cross section is evaluated at the 
energy scale corresponding to the rapidity gap between the 
vector meson and the target $\xpom$.
To translate this into the photon-target center of mass energy $W$  
that is often used to present experimental results note that 
$\xpom = (M_{J/\Psi}^2 + Q^2)/(W^2 + Q^2)$.
The factor $1+\beta^2$ accounts for the
real part of the scattering amplitude and the factor $R_g^2$ corrects
for the skewedness effect, i.e. that the gluons in the target are probed at 
slightly different $x$~\cite{Shuvaev:1999ce,*Martin:1999wb}. 
For these corrections
we follow the prescription of Ref.~\cite{Watt:2007nr}, taking them 
as
\begin{eqnarray}
\beta &=& \tan \frac{\pi \lambda}{2}
\\
R_g &=& \frac{2^{2 \lambda+3}}{\sqrt{\pi}}\frac{\Gamma(\lambda + 5/2)}{\lambda+4} 
\quad \textrm{ with}
\\
\lambda &=& \frac{\partial \ln \A_{T,L}^{\gamma^*p\to J/\Psi p}}{\partial \ln 1/\xpom}.
\end{eqnarray}
These corrections depend, in general, on $t$, which we take into account in our calculation. 
For the full IPsat model $\lambda$ changes by about 5\% between $t=0$ and 
$-t=0.5\gev^2$. For the factorized impact parameter dependence in 
\eqs\nr{eq:factbt} and~\nr{eq:BEKWfact} $\lambda$ is 
independent of $t$.
We calculate the correction terms from the energy dependence of the nucleon
scattering amplitudes and use the same values for the nucleus at the 
same $Q^2,\xpom$. Since the difference in $\lambda$ extracted from the nucleus 
and the nucleon cross sections is small (compared to the value of $\lambda$) and 
$R_g$ and $\beta$ are in themselves corrections to the cross section, this 
approximation is justified. In addition this approximation has the advantage that 
these corrections cancel  on the nucleus/nucleon cross section ratio.
The real part and skewedness corrections, especially $R_g$ are, however, a significant 
factor in the absolute normalization of the cross section and are 
necessary for the agreement with HERA data.

The imaginary part of the scattering amplitude is the Fourier-transform of the
dipole cross section from $\bt$ to $\Deltat$ contracted with the 
overlap between the vector meson and virtual photon wave functions:
\begin{multline}\label{eq:ampli}
\A(\xpom,Q^2,\Deltat) 
= \int \ud^2 \rt \int \frac{\ud z}{4\pi} \int \ud^2 \bt 
\\
\times [\Psi_V^* \Psi](r,Q^2,z)
e^{-i \bt \cdot  \Deltat}  
\dsigma(\bt,\rt,\xpom),
\end{multline}
where we have followed the normalization convention of~\cite{Kowalski:2006hc}.
For the virtual photon--vector meson wavefunction overlap we
use the ``boosted Gaussian'' parametrization from Ref.~\cite{Kowalski:2006hc}.
We have also tested the ``gaus-LC'' wavefunction also used in 
Ref.~\cite{Kowalski:2006hc}. Although the ``boosted Gaussian'' seems preferred
by HERA data, also the ``gaus-LC'' parametrization is compatible with 
the data within the experimental errors. The cross sections for the
proton differ by factors of the order of 10\%.
The interaction of the gluon target with the dipole can in general depend also on 
$\Deltat$, which introduces terms that couple $\rt,$ $\Deltat$  and $z$ in 
\eq\nr{eq:ampli}. For the $J/\Psi$ and the range in $t$ considered in this paper
$\Deltat$ is sufficiently small compared to the relevant values of $1/r$ that 
we can neglect this coupling, which simplifies the structure  considerably.
Lighter vector mesons would require a more general treatment.

The average over the positions of the nucleon in the nucleus is denoted here by
\begin{equation} \label{eq:aavg}
\Aavg{\mathcal{O}(\{ \bt_i \})} 
\equiv \int \prod_{i=1}^{A}\left[ \ud^2 \bt_i T_A(\bt_i) \right] 
\mathcal{O}(\{ \bt_i \}).
\end{equation}
Here $T_A$ is the Woods-Saxon distribution with nuclear radius 
$\ra = (1.12 A^{1/3}-0.86 A^{-1/3})\fm $ and surface thickness 
$d=0.54\fm$.
This expectation value is equivalent to the average
over nucleon configurations in a Monte Carlo Glauber 
calculation.
We are assuming that the positions $\bt_i$ are independent, i.e. 
neglecting nuclear correlations that would be a subject of
interest in their own right (see e.g.~\cite{Alvioli:2009ab}).
The coherent cross section is obtained by averaging the amplitude
before squaring it, $|\Aavg{\A}|^2$, and the incoherent one is
the variance $\Aavg{|\A|^2} -  |\Aavg{\A}|^2$ that measures the fluctuations
of the gluon density inside the nucleus. Because
$\Aavg{\A}$ is a very smooth function of $\bt$, its Fourier transform 
vanishes rapidly for $\Delta \gtrsim 1/\ra$. Therefore at large $\Delta$
the quasielastic cross section \nr{eq:xsec} is almost purely incoherent.

\begin{figure}
\includegraphics[width=0.5\textwidth]{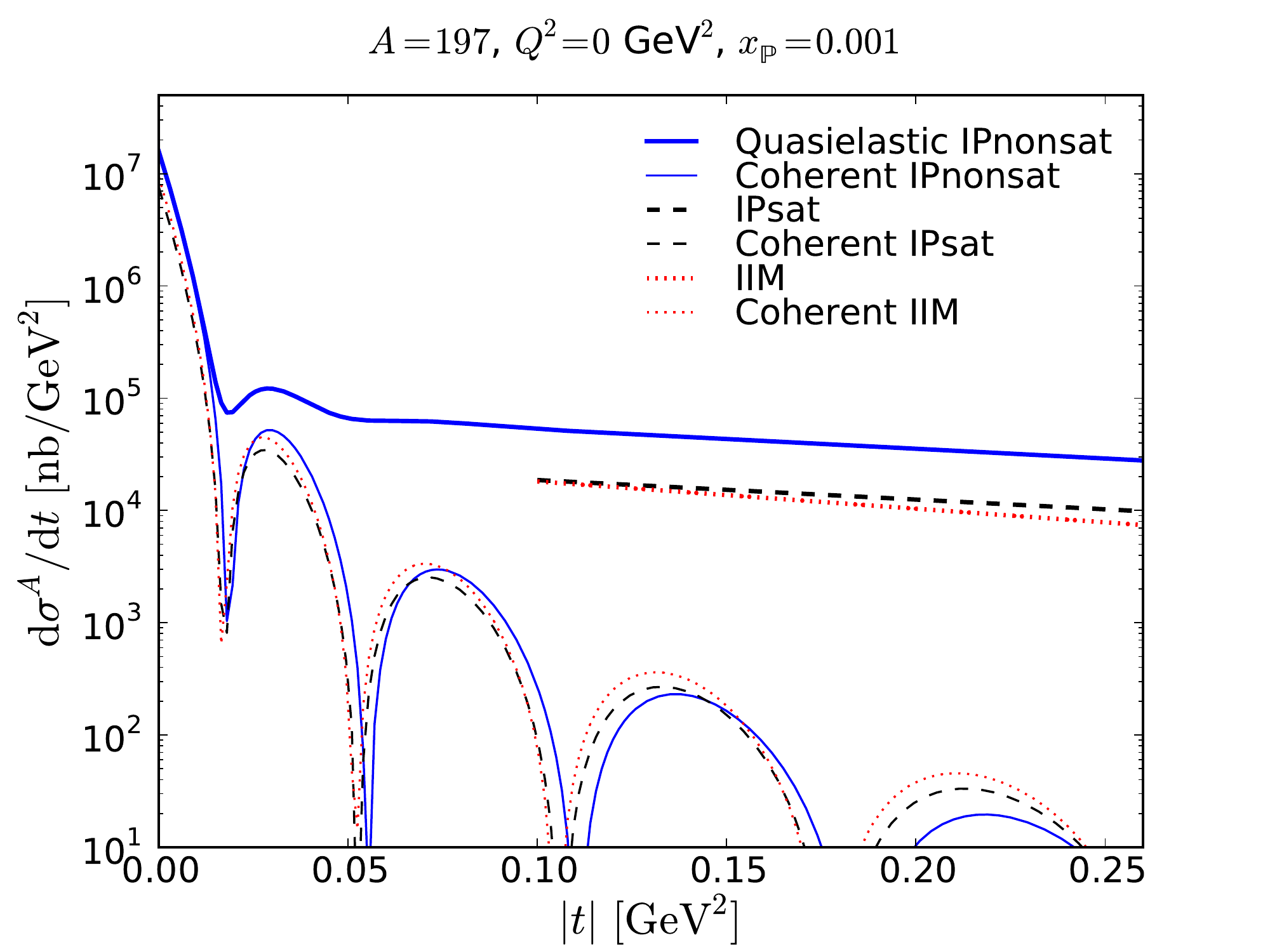}
\caption{The quasielastic and coherent diffractive $J/\Psi$ cross sections in gold
nuclei  at  $Q^2= 0$ and  $\xpom = 0.001$.
Shown are 
the IPsat and IIM parametrizations. We also show the 
result for the linearized  ``IPnonsat'' version 
(used e.g. in Ref.~\cite{Caldwell:2009ke}) where the incoherent
cross section is explicitly $A$ times that of the proton. 
Our approximation \nr{eq:amplisq} 
is not valid for small  $|t|$; the corresponding part of the distribution
has been left out.
} \label{fig:dsigmavst}
\end{figure}

The cross section for quasielastic vector meson production is now expressed
in terms of the dipole scattering amplitude as
\begin{multline} \label{eq:incxsec}
\frac{\ud \sigma^{\gamma^* A \to V A^* }}{\ud t} 
= 
\frac{R_g^2(1+\beta^2)}{16\pi}  
\int 
\frac{\ud z}{4 \pi}  \frac{\ud z'}{4 \pi}
\ud^2 \rt \ud^2 \rt'
\\ \times
\left[ \Psi^*_V \Psi \right] (r,z,Q)
\, \left[ \Psi^*_V  \Psi \right](r',z',Q)
\\ \times
\Aavg{ \left| \mathcal{A}_{q\bar{q}}\right|^2(\xpom,r,r',\Deltat) } \, .
\end{multline}
We now average the square of the dipole scattering amplitude over the 
nucleon coordinates, using the assumptions of
\eqs\nr{eq:sfact} and~\nr{eq:factbt} and taking the large $A$ limit.
We are additionally assuming that $T_A$ is a smooth function on the 
discance scale defined by $B_p$.
Averaging the square of the amplitude gives the total quasielastic 
contribution, but we only keep the terms 
that contribute at large $|t| \gg 1/\ra^2$, which leaves us 
with the expression 
\begin{multline}\label{eq:amplisq}
\left| \mathcal{A}_{q\bar{q}}\right|^2(\xpom,r,r',\Deltat)  
=
16 \pi B_p \int \ud^2 \bt \sum_{n=1}^A
\frac{1}{n} \binom{A}{n} 
\\ \times 
e^{-B_p \Deltat^2/n}
e^{-2 \pi B_p A T_A(b)
\left[ \ampli(r) + \ampli(r') \right] } 
\\ \times
\left( \frac{\pi B_p \ampli(r)\ampli(r') T_A(b) }
  {1 - 2 \pi B_p T_A(b)\left[ \ampli(r) + \ampli(r') \right] } \right)^n
.
\end{multline}
Note that  \eqs\nr{eq:sfact} and~\nr{eq:factbt} have enabled us to
write the leading contributions as proportional to the
(Gaussian) proton impact parameter profile, which can then be 
Fourier-transformed analytically. Giving up either of these approximations
would force us to numerically Fourier-transform the ``lumpy'' 
$b$-dependence corresponding to a fixed configuration
of the nucleon positions. This would make the numerical calculation 
much more demanding and is left for future work. 

The terms with $n \geq 2$ correspond to scattering off a system of several 
overlapping nucleons simultaneously, leading to slower suppresion with $|t|$.
In practice we have verified numerically that they do not
contribute to our results at the values of  $t$ we are interested in 
(the $n=2$ contribution is typically $\lesssim 2\%$ of the $n=1$-one,
only reaching $5\%$ at $-t\gtrsim 0.5 \gev^2$ ) and 
will neglect them in the following. This leaves us with the expression
\begin{multline}\label{eq:amplisqn1}
\left| \mathcal{A}_{q\bar{q}}\right|^2(\xpom,r,r',\Deltat)  
=
16 \pi B_p A \int \ud^2 \bt 
\\ \times 
e^{-B_p \Deltat^2}
e^{-2 \pi B_p A T_A(b)
\left[ \ampli(r) + \ampli(r') \right] } 
\\ \times
\left( \frac{\pi B_p \ampli(r)\ampli(r') T_A(b) }
  {1 - 2 \pi B_p T_A(b)\left[ \ampli(r) + \ampli(r') \right] } \right).
\end{multline}
Equation \nr{eq:amplisqn1} has a very clear interpretation. The 
squared amplitude is proportional to $A$ times the squared amplitude
for scattering off a proton, corresponding to the dipole scattering 
independently off the nucleons in a nucleus. This sum of independent
scatterings is then multiplied by a nuclear attenuation factor
\begin{multline}
\frac{ e^{-2 \pi B_p A T_A(b) \left[ \ampli(r) + \ampli(r') \right] } } 
 {1 - 2 \pi B_p T_A(b)\left[ \ampli(r) + \ampli(r') \right] } 
\approx
\\
 e^{-2 \pi (A-1) B_p T_A(b) \left[ \ampli(r) + \ampli(r') \right] } ,
\end{multline}
which accounts for the requirement that the dipole must \emph{not}
scatter inelastically off the other $A-1$ nucleons in the target (otherwise the
interaction would not be diffractive). 
Note that the factor 
$4 \pi B_p \ampli(r,\xpom)=\sigmap(r,\xpom)$ is the proton-dipole cross section for  a
dipole of size $r$. Thus this attenuation corresponds to the
probability of a dipole with a cross section which is the average 
of dipoles with $r$ and $r'$ to pass though the nucleus.
A similar expression 
can be found e.g. in Ref.~\cite{Kopeliovich:2001xj}.

For comparison, the coherent cross section in our approximation is given by
\begin{equation} \label{eq:coh}
\frac{\ud \sigma^{\gamma^* A \to V A }}{\ud t} 
=\frac{R_g^2(1+\beta^2)}{16\pi} \left| \Aavg{\A(\xpom,Q^2,\Deltat)} \right|^2,
\end{equation}
where in the large $A$ and smooth nucleus limit the amplitude is
\begin{multline}\label{eq:cohampli}
\Aavg{\A(\xpom,Q^2,\Deltat) }
= \int \frac{\ud z}{4\pi} \ud^2 \rt  \ud^2 \bt e^{-i \bt \cdot  \Deltat}  
\\
 \times [\Psi_V^*\Psi](r,Q^2,z)
\,  2 \left[ 1-\exp\left\{ - 2 \pi B_p A T_A(b) \ampli(r,\xpom) \right\} \right].
\end{multline}

\begin{figure}
\includegraphics[width=0.5\textwidth]{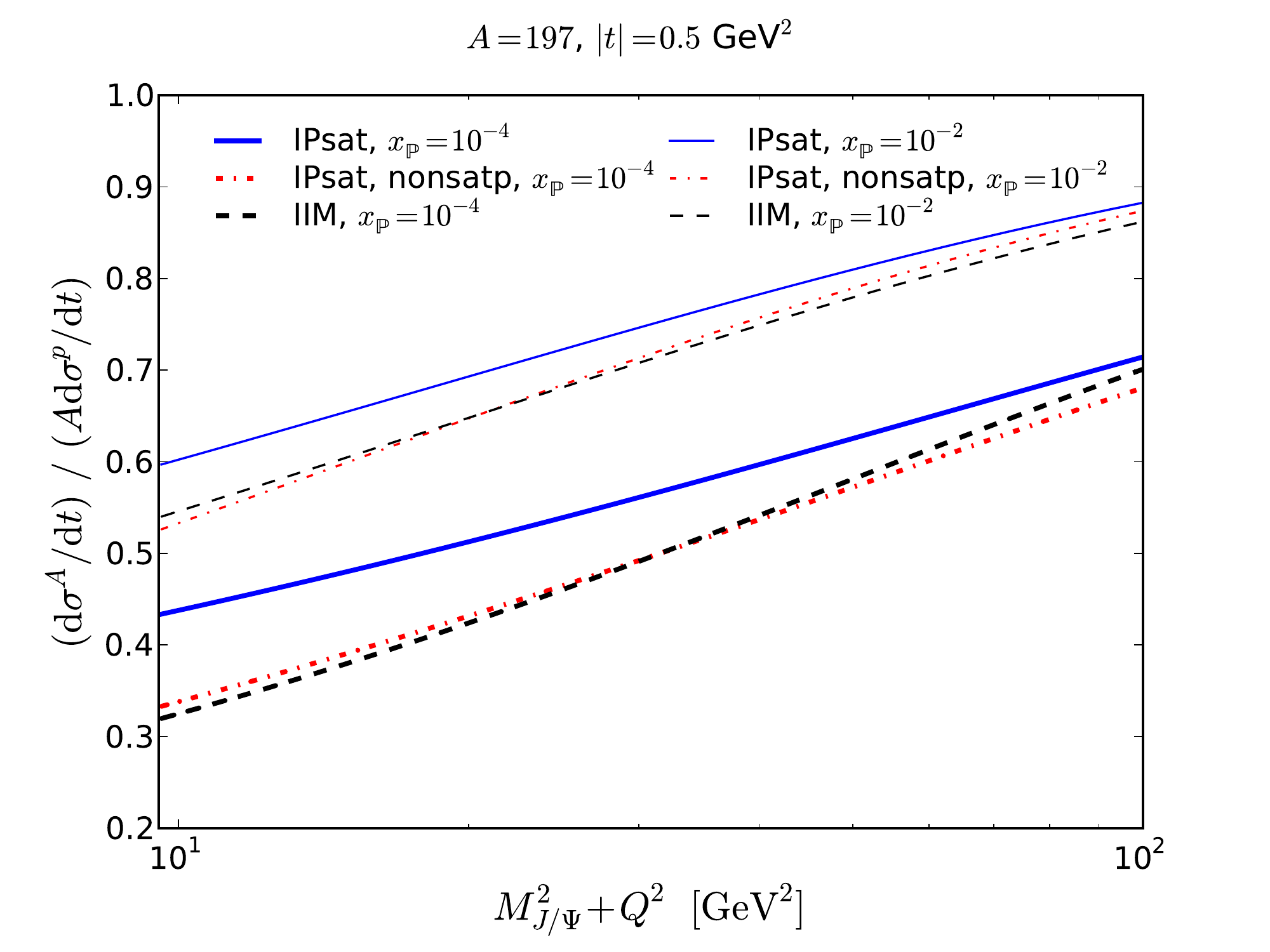}
\caption{The ``nuclear transparency'' 
ratio of cross sections vs. $Q^2$ for IPsat, IIM 
parametrizations at $\xpom= 10^{-2}$ (the upper three curves, blue)
and $10^{-4}$ (the lower 3 curves, black).
For comparison we also include
we also include the result if unitarization effects are included 
at the nucleus but not at the nucleon level in the IPsat-parametrization. 
(See text for discussion).
}\label{fig:ratiovsq}
\end{figure}

\begin{figure}
\includegraphics[width=0.5\textwidth]{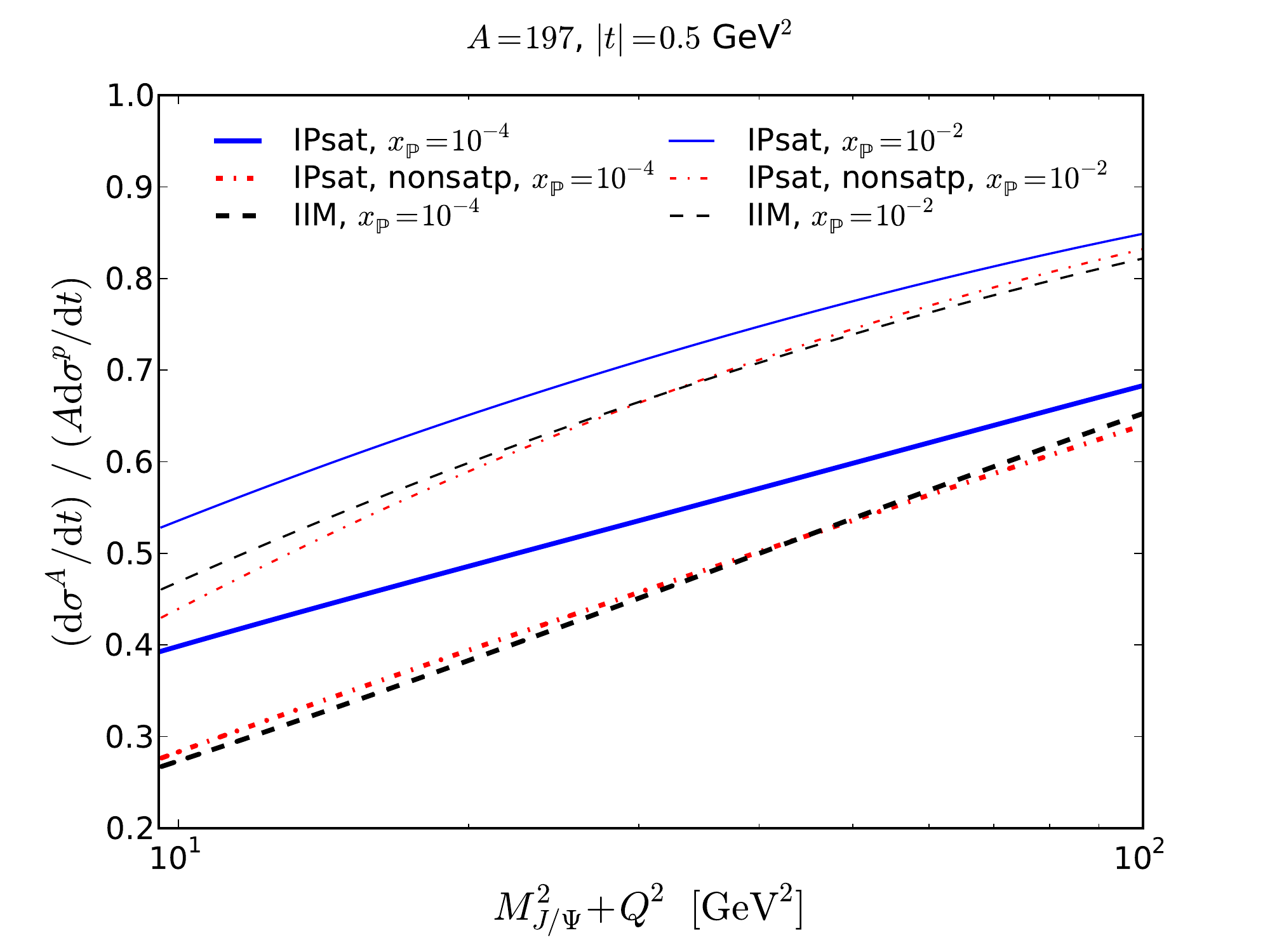}
\caption{The ``nuclear transparency'' 
ratio of cross sections vs. $Q^2$ using the ``Gaus-LC'' vector meson 
wavefunctions. The labeling is the same as in \fig\ref{fig:ratiovsq}.
}\label{fig:ratiovsq_gauslc}
\end{figure}

\section{Results and discussion}
\label{sec:res}

We first test our dipole cross section parametrizations and vector meson wave
functions by comparing them to HERA results~\cite{Chekanov:2004mw,*Aktas:2005xu} 
on diffractive $J/\Psi$ production
that is known to be well described by dipole model 
fits~\cite{Kowalski:2006hc,Marquet:2007qa}. The comparison is quite satisfactory, as
can be seen from \fig\ref{fig:hera}.  In addition to the factorized 
approximation (\eq\nr{eq:BEKWfact}, ``factorized IPsat'' in the figure) that 
we are using in the rest of this paper, also shown is the result with the original
IPsat parametrization (\eq\nr{eq:unfactbt}, denoted ``IPsat'' in the figure). The 
factorized approximation differs from the original one slightly at small 
$Q^2$, but the difference is not significant for our purpose of establishing a 
reasonable baseline for computing nuclear effects.

We note here that the diffractive slope parameters
in the parametrizations are different, $B_p=4.0\gev^{-2}$ for IPsat and 
$B_p= 5.59\gev^{-2}$
for IIM; since these are correlated with the other parameters in the fits leading to
the parameter values used we do not wish to alter them here.
Our approximation of a factorized $b$-dependence with a constant $B$ 
does not allow us to describe the observed weak energy and $Q^2$ dependence of the
diffractive slope. 
The larger $B$ that we use for IIM comes from the $\sigma_0$ normalization
in a fit to inclusive $F_2$ data, and also agrees with the observed slopes in 
inclusive diffraction at large $\beta$ and small 
$\xpom$~\cite{Chekanov:2004hy,*Aktas:2006hx} and exclusive $\rho$ and $\phi$
data~\cite{Adloff:1999kg,*Chekanov:2005cqa}. The HERA $J/\Psi$-data,
on the other hand, has a smaller slope 
$\sim 4\gev^{-2}$~\cite{Chekanov:2004mw,*Aktas:2005xu}. 
The $t$-slope in
the IPsat parametrization is mostly determined by this $J/\Psi$-measurement,
and an agreement with the larger measured slopes for $\rho$ and $\phi$
is obtained by taking into account the larger size of the wavefunctions
of these lighter mesons.

\begin{figure}
\includegraphics[width=0.5\textwidth]{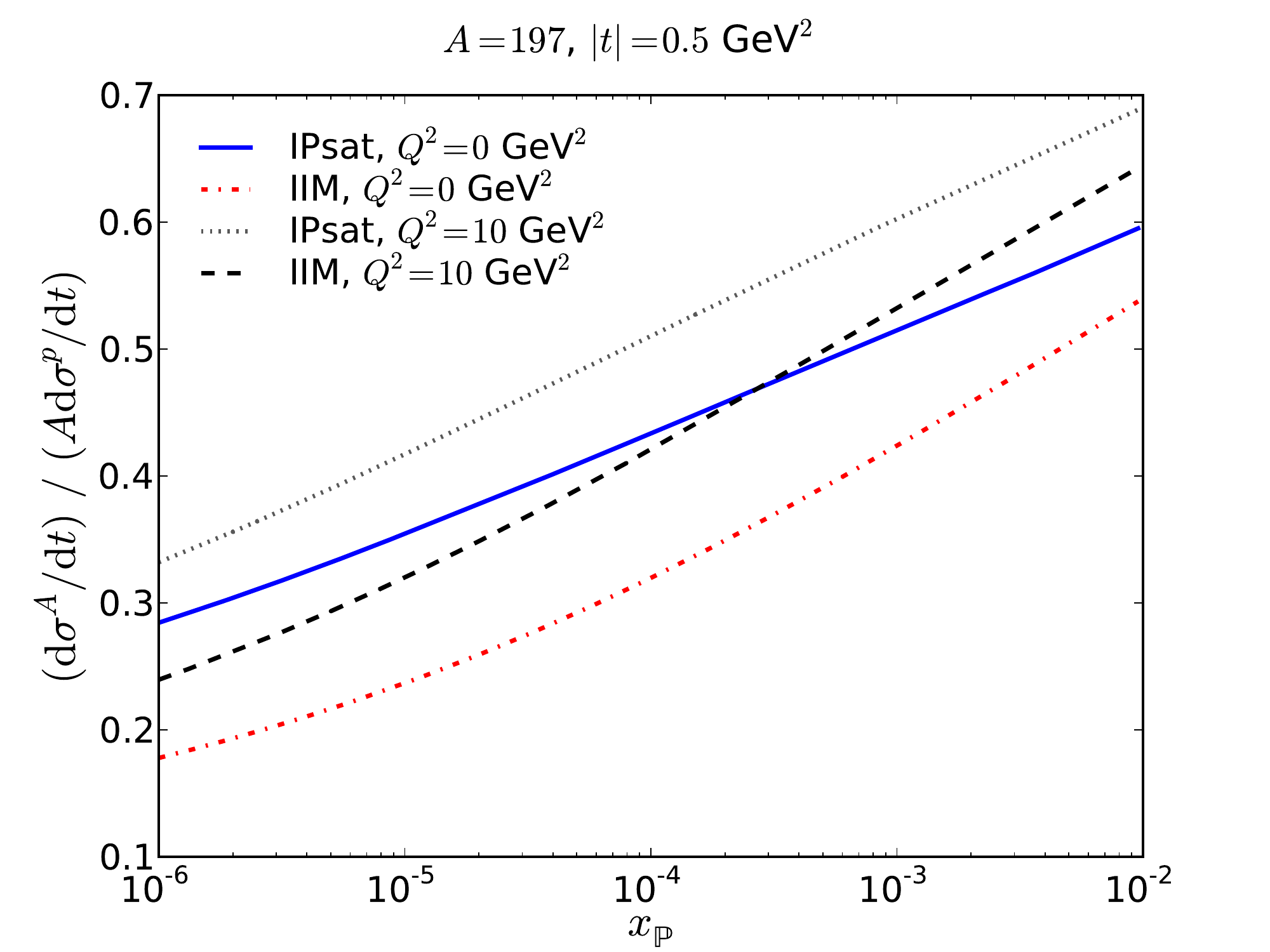}
\caption{The ``nuclear transparency''  ratio of cross sections vs. $\xpom$ 
using the IPsat and IIM  parametrizations for $Q^2=0$ and $Q^2=10\gev^2$.
}\label{fig:ratiovsx}
\end{figure}

The differential cross section $\ud \sigma^{\gamma^* A \to J/\Psi A}/\ud t$
 for $A=197$ (gold) as a function of $t$ is presented in \fig\ref{fig:dsigmavst}. 
We show the cross 
section at $\xpom = 0.001$ 
for photoproduction.  As we performed
the nuclear wavefunction average leading to  \eq\nr{eq:amplisq} in the 
approximation where $|t|$ is large, neglecting the coherent contribution,
we cannot extend our incoherent curves to small $|t|$. 
For comparison we show the corresponding
``IPnonsat'' result where the IPsat model is linearized in $r^2 F(x,r)$. 
This curve corresponds to the calculation done in Ref.~\cite{Caldwell:2009ke},
including both the coherent and incoherent contributions, but without the effect
of multiple scattering off different nucleons (i.e. the incoherent cross section is
explicitly $A$ times the one for a proton). 
As one can see, the nuclear modification
due to multiple scattering (resulting mostly from the
factor $e^{-2 \pi B_p  A T_A(b) \left[ \ampli(r) + \ampli(r') \right] } $  in 
\eq\nr{eq:amplisq}) is very large. 
 In the full black disk limit 
of $\ampli(r)=1$ this factor becomes $\approx e^{-0.5 A^{1/3}}$ and completely
supresses the contribution from the center of a large nucleus, leaving only an area
of $\approx 2 \pi d \ra \sim A^{1/3}$ contributing to the integral over $\bt$.
Thus the cross section in 
the black disc limit behaves as $\sim A^{1/3}$ compared to $\sim A$ in the dilute
limit, so a large suppression is to be expected.

We also show in \fig\ref{fig:dsigmavst} the coherent cross sections 
(using \eq\nr{eq:cohampli}). They are also suppressed compared to the linearized 
version (IPnonsat), but not by as much as the incoherent one. In the linearized
version (as can be seen explicitly in Ref.~\cite{Caldwell:2009ke} where this case
was considered) the ratio between the coherent cross section at $t=0$ and the incoherent
one extrapolated to $t=0$ is $A$. In the IPsat model we get
$270$ ($250$) and in the IIM model $300$ ($270$) at $Q^2=0$ ($Q^2=10\gev^2$).
This would make it slightly easier to  measure the first 
diffractive dip in the coherent cross section, since the background from the incoherent
process is smaller by a factor of 2 than the linearized 
estimate~\cite{Caldwell:2009ke}.

\begin{figure}
\includegraphics[width=0.5\textwidth]{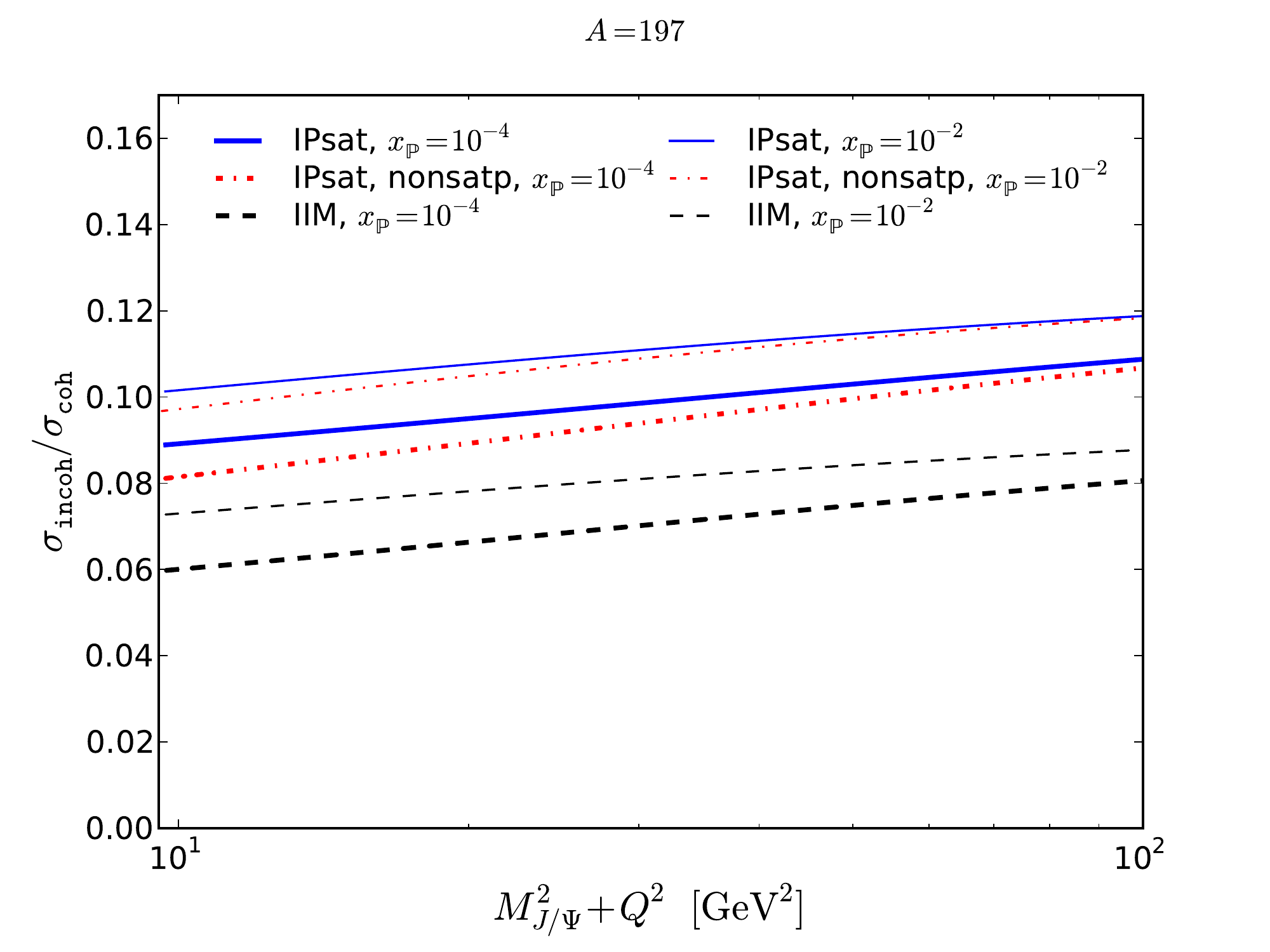}
\caption{
The incoherent cross section integrated over the interval $0.1\gev^2< -t < 0.3 \gev^2$
divided by the coherent cross section integrated over $0 < -t < 0.1 \gev^2$
as a function of $Q^2 + M_{J/\Psi}^2$.
} \label{fig:incohovercohQ}
\end{figure}

\begin{figure}
\includegraphics[width=0.5\textwidth]{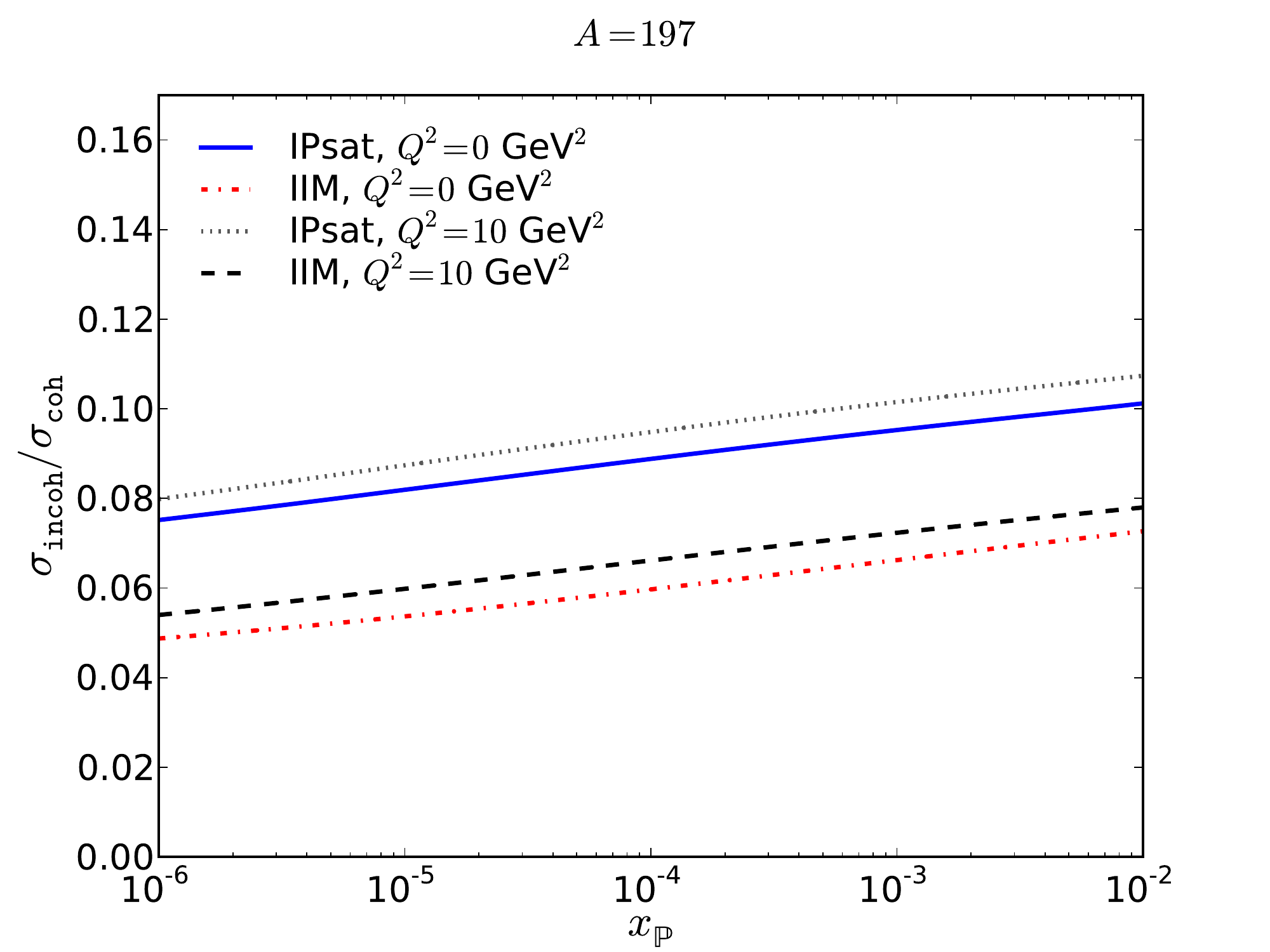}
\caption{
The incoherent cross section integrated over the interval $0.1\gev^2< -t < 0.3 \gev^2$
divided by the coherent cross section integrated over $0 < -t < 0.1 \gev^2$
as a function of $\xpom$.
} \label{fig:incohovercohx}
\end{figure}

To demonstrate the nuclear dependence further we show in \fig\ref{fig:ratiovsq}
the ratio of the cross section in a gold  nucleus to that in a nucleon as a 
function of $Q^2$. Historically this ratio is known as the ``nuclear transparency''.
Its smallness at low energy, similarly to coresponding quantities in 
hadron-nucleus scattering, is due to the interactions of the $J/\Psi$ as it propagates
through the nucleus. The growth of the transparency towards $1$ for increasing
$Q^2$~\cite{Adams:1994bw,Arneodo:1994qb,*Arneodo:1994id} is 
a demonstration of \emph{color transparency} (see e.g. 
Ref.~\cite{Frankfurt:1991nx,Frankfurt:1993it,Brodsky:1994kf,Kopeliovich:2001xj,Frankfurt:2005mc,Miller:2010eh}),
namely that at large
$Q^2$ the interacting components of the photon wavefunction are of smaller
size $r$ and interact weakly. In our framework color transparency is
automatically present in the fact that the dipole cross section approaches zero
for $r\to0$. In \fig\ref{fig:ratiovsq} we also show the result (labeled
``IPsat, nonsatp'') of using a nonsaturated dipole-nucleon cross section
in \eq\nr{eq:amplisq}. This corresponds to including 
unitarity effects at the nucleus level but not 
for a single nucleon. The observed nuclear suppression in this unphysical 
scenario is significantly larger than for the saturated full IPsat
parametrization, showing the sensitivity of the nuclear transparency
to saturation effects already at the proton level.

The IIM parametrization has a much larger nuclear 
suppression in incoherent diffraction, with the nuclear transparency
ratio close that of an unsaturated dipole-proton cross section.
To put this in perspective recall that both parametrizations
gave an equally good description of the elastic cross section
measured at HERA (\fig\ref{fig:hera}). Since IIM does this with a
larger $B_p$ than IPsat, we can infer that the typical $\ampli$ is smaller, 
so that the elastic cross section  $\sigma^\textrm{el}
\sim B_p \ampli^2$ is of the same order. 
The nuclear transparency ratio, on the other hand, depends on the
total dipole-nucleon cross section 
$\sim B_p \ampli \sim \sigma^\textrm{el}/\ampli$ which is thus 
larger for IIM. Thus we have a situation where both parametrizations
have been fitted to inclusive $F_2$ data\footnote{Although we have here 
approximated the original IPsat parametrization by factorizing the $b$-dependence.},
reproduce well the HERA $J/\Psi$ cross section, but differ in their
result for incoherent diffraction in nuclei. This stresses the importance of
performing a global analysis of both inclusive and diffractive data
to constrain the dipole cross sections, and demonstrates
the utility of eventual incoherent diffractive measurements in such 
an analysis.

Figure \ref{fig:ratiovsq_gauslc} shows the same $Q^2$-dependence
using the ``gaus-LC'' wavefunction. It puts more
weight on large dipole sizes, leading to a stronger nuclear suppression. 
The cross section ratio typically decreases by $\sim 0.04$ from 
the ``boosted Gaussian'' wavefunction, but the
relative structure between the different dipole cross sections stays the same.
The difference between the cross sections themselves is larger,
but much of the it cancels in the ratio. The existing HERA data 
is not precise enough to fully discriminate between different models for
the vector meson wavefunction, a situation which should also improve
with planned new DIS experiments.

\begin{figure}
\includegraphics[width=0.5\textwidth]{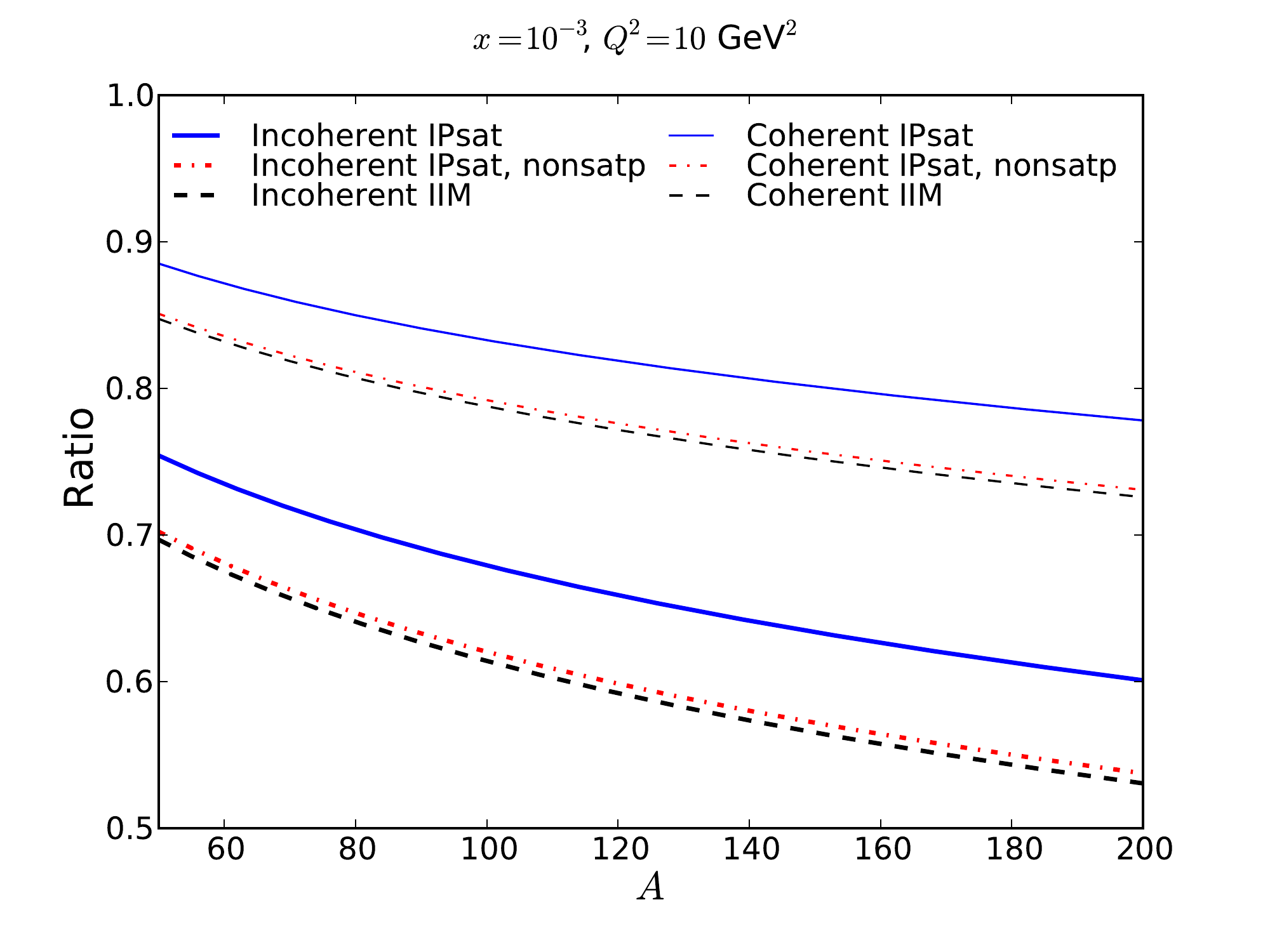}
\caption{
The ratio of the coherent (at $t=0$) and incoherent (at $t=-0.5\gev^2$,
but in our approximation this does not depend on $t$) 
cross sections to the corresponding ones for a proton; normalized with 
$A^2$ and $A$ respectively. Plotted as a function of $A$,
for $\xpom = 0.001$ and $Q^2=10\gev^2$.
} \label{fig:cohincoh}
\end{figure}

The energy dependence of the nuclear suppression (again for $A=197$)
is shown in \fig\ref{fig:ratiovsx} for both IPsat and IIM parametrizations 
at $Q^2=0$ and $Q^2=10\gev^2$. Again 
we see the larger nuclear suppression in the IIM model than in IPsat.
The differences in the energy (i.e. $\xpom$) dependence of
the two dipole cross sections are more clearly visible in the 
photoproduction result. This is natural, since in the IPsat model 
the energy dependence at the initial scale of the DGLAP evolution
(probed at smaller $Q^2$) is almost flat, in stark contrast to the
typical behavior resulting from BK evolution. At higher $Q^2$
the difference in the $x$-dependence is smaller, although 
there the IPsat-model, driven by the DGLAP evolution, 
turns over to a \emph{faster}
energy dependence. We have not extrapolated our curves to higher 
energies, since there is no prospect of experimental measurements. One does 
however see from \fig\ref{fig:ratiovsx} that the curves
continue to go down when extrapolated to smaller $\xpom$. This is to be 
expected since, as discussed previously, one has not yet reached the  
black disk limit. 

In a realistic experimental setup it might be possible to detect 
or veto the nuclear breakup even when the momentum transfer
$t$ is not measured very accurately. In this case it will be interesting 
to understand how the relative magnitudes of the incoherent and coherent
cross sections behave as a function of $Q^2$ and $\xpom$. Generally when approaching 
the black disk limit the coherent cross section increases and the incoherent one
decreases. The relative change shows, however, a smaller dependence on $Q^2$ and 
$\xpom$ than the nucleus/nucleon cross section ratio. This is shown in our 
parametrization in  \figs\ref{fig:incohovercohQ} and~\ref{fig:incohovercohx}, where
we plot the 
the incoherent cross section integrated over the interval $0.1\gev^2< -t < 0.3 \gev^2$
divided by the coherent cross section integrated over $0 < -t < 0.1 \gev^2$
as a function of $Q^2 + M_{J/\Psi}^2$ and $\xpom$.
Figure \ref{fig:cohincoh} further demonstrates the relative similarity of the 
nuclear suppression in the coherent and incoherent cross sections. 
Shown is the $A$ dependence of the ratios 
$ (\ud\sigma^A_\mathrm{incoh}/\ud t)/(A \ud \sigma^p/\ud t)$
(which, in our approximation, is independent of $t$)
and 
$\left. (\ud\sigma^A_\mathrm{coh}/\ud t)/(A^2 \ud \sigma^p/\ud t)\right|_{t=0}$
for $Q^2=10\gev^2$ and $\xpom=0.001$. Note that the coherent and the incoherent 
cross sections are normalized by different powers of $A$ and that
width of the coherent peak at small $t$ also depends on $A$.

Figures \ref{fig:dsigmavst} and \ref{fig:ratiovsq} are our main result.
Our calculation uses as input only well tested parametrizations that have been fit
to existing HERA data and nuclear geometry. We work strictly in the
small $x$-limit which makes our formalism simple and transparent.
This paper provides realistic estimates for the absolute cross sections 
that could be measured in future nuclear DIS experiments. 
We have, however, made several simplifying assumptions in our calculation, the most 
important being a) the factorized impact parameter dependence \eq\nr{eq:factbt},
b) the assumption of independent scattering off different nucleons
\eq\nr{eq:sfact} and c) neglecting nucleon-nucleon correlations. Including
these effects in a physically correct manner and discussing how they could be 
studied experimentally is left for future work. As can be seen from 
the values of the nuclear suppression in \figs\ref{fig:ratiovsq} and~\ref{fig:ratiovsx},
the effects of high  densities, gluon saturation and unitarity on the
incoherent cross section are large.
Thus incoherent diffraction in future nuclear DIS experiments will
be a sensitive probe of small-$x$ physics.

\section*{Acknowledgements}
We thank E. Aschenauer, H. Kowalski and W. Horowitz for discussions and K.~J.~Eskola
for a careful reading of the manuscript.
The work of T.L. has been supported by the Academy of Finland, project 
126604. T.L. wishes to thank the INT  at the 
University of Washington for 
its hospitality during the completion of this work.

\bibliography{spires}
\bibliographystyle{JHEP-2modM}

\end{document}